\documentclass{aa} 
\usepackage{amsmath,amssymb}
\usepackage{bm}
\usepackage{threeparttable}
\usepackage[varg]{txfonts}
\usepackage{natbib}
\bibpunct{(}{)}{;}{a}{}{,} 
\usepackage{url}
\usepackage{graphicx}
\usepackage[colorlinks=true, linkcolor=blue, citecolor=blue, filecolor=blue, urlcolor=blue]{hyperref}

\begin{document} 
\title{Density fluctuation in the solar corona and solar wind:\\
A comparative analysis of radio-occultation observations \\ and magnetohydrodynamic simulation}
\titlerunning{Density fluctuation in the solar corona and solar wind}


\author{Shota Chiba\inst{\ref{inst1}, \ref{inst2}}
            \and
        Munehito Shoda\inst{\ref{inst3}}
            \and
        Takeshi Imamura\inst{\ref{inst1}}
        }
        
\institute{
                Graduate School of Frontier Sciences, 
                The University of Tokyo, 5-1-5 Kashiwanoha, Chiba, 277-8561, Japan\label{inst1}
                \and
                Nagoya University, Furo-cho, Chikusa-ku, Nagoya, 464-8601, Japan \label{inst2}
                \and
                Department of Earth and Planetary Science, School of Science, The University of Tokyo, \\
                7-3-1 Hongo, Bunkyo, Tokyo, 113-0033, Japan\label{inst3}
                }           
\date{}

   \abstract
   {
   Recent in situ observations and numerical models indicate that various types of magnetohydrodynamic (MHD) waves contribute to the solar wind acceleration. 
   Among them is an MHD wave decomposition at distances closer than 50~$R_{\sun}$ using data taken by the first perihelion pass of \textit{Parker Solar Probe} (PSP).
   However, the underlying physical processes responsible for the formation of the solar wind have not yet been observationally confirmed at distances closer than 10~$R_{\sun}$.
   }
   {
   We aim to infer the mode population of density fluctuations observed by radio occultation, which has all been attributed to slow magnetoacoustic waves.
   }
   {
   We compare the radio occultation observations conducted in 2016 using the JAXA’s Venus orbiter \textit{Akatsuki} with the MHD simulation. The time-frequency analysis was applied to the density fluctuations observed by the radio occultation and those reproduced in the MHD model.
   }
   {
   The time-spatial spectrum of the density fluctuation in the model exhibits two components that are considered to be fast and slow magnetoacoustic waves. 
   The fast magnetoacoustic waves in the model tend to have periods shorter than the slow magnetoacoustic waves, and the superposition of these modes has a broadened spectrum extending in the range of approximately 20–1000 s, which resembles that of the observed waves. 
   }
   {Based on this comparison, it is probable that the density oscillations observed by radio occultation include fast and slow magnetoacoustic waves, and that fast magnetoacoustic waves are predominant at short periods and slow magnetoacoustic waves are prevalent at long periods. 
   This is qualitatively similar to the results of the mode decomposition obtained from the PSP's first perihelion at more distance regions.}
   
   \maketitle
%
\section{Introduction}\label{sec:intro}
The mechanism for solar wind acceleration is one of the most crucial problems in solar and space plasma physics, and the coronal heating problem is inseparable from this acceleration \citep{Cranmer2019}.
The acceleration mainly occurs in the outer corona at heliocentric distances of about 2--10~$R_{\sun}$ (= solar radii) \citep[e.g.,][]{Armstrong1981, Muhleman1981, Scott1983, Coles1995, Tokumaru1991, Tokumaru1995}, where coronal heating by magnetohydrodynamic (MHD) waves and the wave-induced magnetic pressure are thought to play major roles. The wave- and turbulence-driven model \citep{Cranmer2012} stands out as a notable framework for understanding solar wind acceleration, proposing acceleration by Alfv\'{e}n waves originating on the solar surface \citep{Suzuki2005, Matsumoto2012, Cranmer2007, vanderHolst2014, Usmanov2018, Shoda2019}, although the reconnection/loop-opening processes have also attracted significant recent attention in the literature \citep{Wang2020, Bale2023, Raouafi2023, Iijima2023}.
Within the context of wave- and turbulence-driven wind models, Alfv\'{e}n-wave turbulence, initiated by the interaction of bidirectional Alfv\'{e}n waves, is posited as the primary heating mechanism \citep{Cranmer2007, Verdini2010, Chandran2019}. 
Central to this premise is the concept of wave reflection \citep{Matthaeus1999, Dmitruk2002}. Recent investigations highlight the importance of density fluctuations in the solar wind for effective wave reflection\citep{vanBallegooijen2016, Shoda2018, Shoda2019, Cranmer2023}.
Notably, these density fluctuations are inherently produced by Alfv\'{e}n waves \citep{Suzuki2005, Matsumoto2012, Shoda2019}, underscoring the pivotal role of density fluctuations in the behavior of Alfv\'{e}n waves within the solar wind. Outwardly propagating Alfv\'{e}n waves have been observed in coronagraph images in the lower corona \citep[e.g.,][]{Tomczyk2007}; however, the underlying physical processes responsible for the acceleration of the solar wind have not yet been observationally detected since optical methods have not been effective, and no spacecraft has ever reached there.
\par
Recently, an inner heliosphere observation network composed of NASA's \textit{Parker Solar Probe} (PSP), ESA's \textit{Solar Orbiter} (SolO), and \textit{BepiColombo} is being developed. In particular, PSP will approach as close as 9~$R_{\sun}$ to the Sun and directly explore the solar wind acceleration region. The high temporal resolution data of the density, velocity, temperature, and magnetic field are available from the instruments on board PSP: SWEAP \citep{Kasper2016} and FIELDS \citep{Bale2016}. The measurements during the first perihelion pass of PSP from 53 into 35~$R_{\sun}$ yielded important results, including the ubiquitous presence of the magnetic switchbacks \citep[e.g.,][]{Bale2019, Kasper2019, Dudok2020, Mozer2020}.
\par
\citet{Chaston2020} applied the mode decomposition analysis to PSP data in the solar wind using the technique developed by \citet{Glassmeier1995} and derived MHD wave composition at distances closer than 50~$R_{\sun}$. 
In this method, the composition of the spectral energy density was calculated from the eigenvector describing the MHD variables for each mode. 
According to their results, outwardly propagating Alfv\'{e}n waves were superior to the other modes over the frequency range of 0.000\,2--0.1 Hz, and the contribution from outwardly propagating slow magnetoacoustic waves also had a significant fraction below 0.01 Hz, providing up to 30\% of the spectral composition. 
Although fast magnetoacoustic waves constituted less than 20\%, 
the fraction of these modes increased above 0.01 Hz.
\par
in situ observations can obtain the local characteristics of the solar wind with high temporal resolutions as \citet{Chaston2020} has demonstrated. 
On the other hand, radio occultation observations can cover approximately the coronal base to 20 $R_\sun$ along the heliocentric distance. 
MHD waves in the corona have been investigated by radio occultation observations using spacecraft signals as well as by coronagraph imaging and in situ measurements. 
Observations of Faraday rotations using the Helios and Messenger spacecraft radio occultation signals detected periodical oscillations of the coronal magnetic fields at distances of 1.6--12~$R_{\sun}$, 
which are considered manifestations of Alfv\'{e}n waves \citep{Hollweg1982, Chashei1999, Efimov2015, Jensen2013a, Wexler2017}. 
The observed Alfv\'{e}n waves appear to be in a regime of free propagation based on the radial dependence of the Faraday rotation amplitude \citep{Andreev1997}. 
\citet{Efimov2010, Efimov2012} studied quasi-periodic electron density fluctuations, termed quasi-periodic components (QPCs),
with spacecraft radio occultation observations and identified them as acoustic waves (slow magnetoacoustic waves). 
The typical periods of the QPCs of about three to four minutes are longer than the cut-off period of acoustic waves below the transition region \citep{Erdelyi2007}, 
and thus, the waves could not have propagated from the photosphere. 
JAXA’s Venus orbiter \textit{Akatsuki} has been used for radio occultation observations of the solar corona repeatedly at solar superior conjunctions \citep{Imamura2014}. 
In the observation campaigns conducted in 2011 and 2016, 
QPCs with periods longer than 150 s were ubiquitously observed at 1.5--20.5~$R_{\sun}$ \citep{Miyamoto2014, Chiba2022}. 
They also derived the radial dependences of the wave amplitude, the period, 
the coherence time, and the energy flux to conclude that slow magnetoacoustic waves originate from Alfv\'{e}n waves and are thermalized at distances $\ga$6~$R_{\sun}$.
\par
In this study, we perform a comparative analysis of the quasi-periodic density fluctuation between radio occultations and an MHD simulation, with a view to constraining the physical characteristics of the observed fluctuations. 
The primary objective of this study is to constrain the mode population of disturbances in the solar wind by evaluating the validity of the analysis results by \citet{Chaston2020} from the standpoint of radio occultation observations. 
The remainder of this paper is organized as follows. Section 2 presents the measurement procedure and the analyses of the observational data, Section 3 presents the analyses of the simulation data, Section 4 shows the comparison of the radio occultation data with the MHD simulation, and Section 5 gives the summary.

\section{Observations}\label{Sec_obs}
\subsection{Dataset}\label{Sec_dataset}
\begin{figure}
\resizebox{\hsize}{!}
   {
   \includegraphics{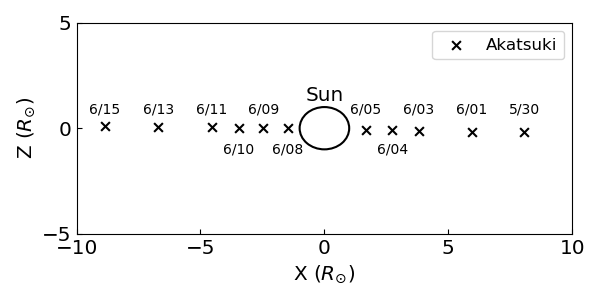}
   }
   \centering
   \caption{Location of Akatsuki (crosses) relative to the Sun (open circle) as seen from the Earth on the date of observations in 2016. 
   The Z axis is directed to the North of the Sun. The numerals indicate the observation dates [M/DD], and the locations are those at 04:45 UT on each day
            }\label{fig_geometory}%
\end{figure}

The observations of the corona and solar wind were conducted 
from May 30 to June 15 2016 using the 8.4 GHz (X-band) downlink signal transmitted by \textit{Akatsuki} and received at the Usuda Deep Space Center (UDSC) of Japan during the superior conjunction of Venus \citet{Chiba2022}. 
The frequency of the transmitted radio wave was stabilized by an onboard ultra-stable oscillator (USO), and the received signal was recorded by an open-loop recording system. 
The solar activity in 2016 was in the intermediate phase between the solar maximum to the solar minimum of Cycle 24. 
The total of 11 observations covered heliocentric distances of 1.36--9.00~$R_{\sun}$ (Figure \ref{fig_geometory}).
The first half period (May 30 to June 5) covered the western side of the Sun, and the second half period (June 8 to 15) covered the eastern side. Table\ref{tab_observations} summarizes the observation conditions for the data used in this study. 
The data are the same as those used in \citep{Chiba2022}, although this study extends the wave analysis to shorter periods by analyzing phase time series retrieved with a higher temporal resolution. The data taken at the minimum heliocentric distance on June 8 was not used since a high noise level prevented the analysis. The details of the radio science subsystem are given in \citep{Imamura2011, Imamura2017}, and the analysis procedure is given in \citep{Chiba2022}.
\begin{table*}
    \caption{Summary of the observations.}
    \label{tab_observations}
    \centering
    \begin{tabular}{l| c c c c c}
        \hline\hline         
        Date & Start time & Length & 
        Heliocentric & Heliocentric & East/West\tablefootmark{***} \\
        &  &  &
        distance\tablefootmark{*} 
        & latitude\tablefootmark{**} &   \\
        &  & [hour] &
            [$R_{\sun}$] & [deg] &   \\
        \hline
        2016-May-30 & 02:00 & 5.5 & 8.19--7.95 & -1.37 & West \\
        2016-June-1 & 02:00 & 5.5 & 6.08--5.83 & -1.52 & West \\
        2016-June-3 & 02:00 & 5.5 & 3.94--3.69 & -1.78 & West \\
        2016-June-4 & 02:00 & 5.5 & 2.87--2.62 & -2.03 & West \\
        2016-June-5 & 02:00 & 5.5 & 1.80--1.56 & -2.58 & West \\
        2016-June-9 & 02:00 & 5.5 & 2.36--2.56 & 0.57 & East \\
        2016-June-10 & 02:00 & 5.5 & 3.33--3.58 & 0.40 & East \\
        2016-June-11 & 02:00 & 5.5 & 4.41--4.61 & 0.46 & East \\
        2016-June-13 & 02:00 & 5.5 & 6.59--6.84 & 0.61 & East \\
        2016-June-15 & 02:00 & 5.5 & 8.75--9.00 & 0.76 & East \\
            \hline
         \end{tabular}
         \tablefoot{\\
   \tablefoottext{*}{Heliocentric distance between Sun's center and the tangential point of the ray path.}
   \\
   \tablefoottext{**}{Heliographic latitude at 04:45 on each day.}\\
   \tablefoottext{***}{Eastern or western side of the Sun as seen from the Earth.}
   }
\end{table*}
\begin{figure}[htpb]
   \resizebox{\hsize}{!}{\includegraphics{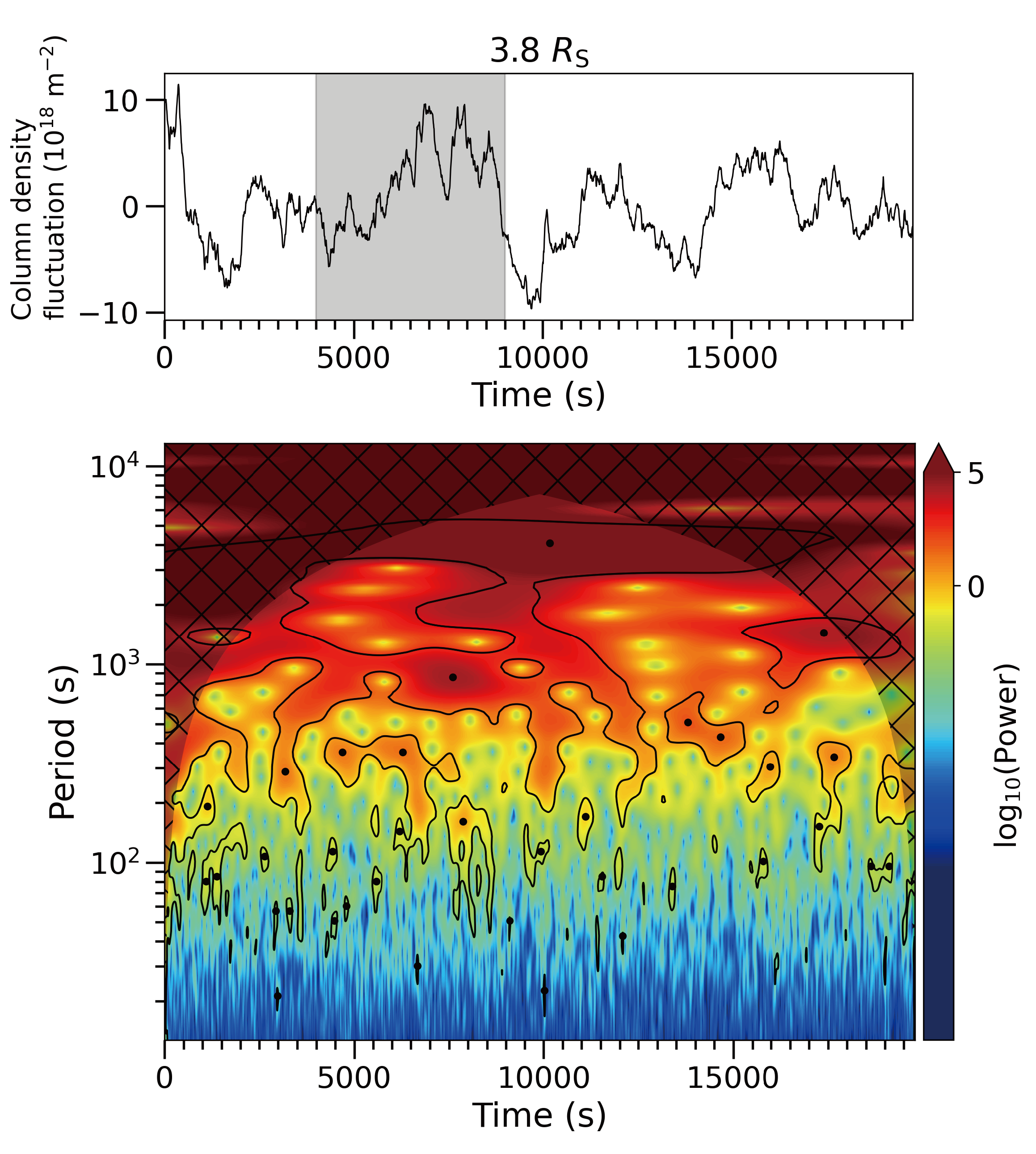}
   }
   \centering
   \caption{Example of (upper) the time series of the electron column density fluctuation and (lower) its wavelet spectrum.
   The data was obtained on June 3, 2016, at a heliocentric distance of 3.8~$R_{\sun}$. 
   The wavelet power is normalized by the variance of the time series. 
   The shaded region in the lower panel is the cone of influence (COI), 
   which is the region where the effect of the discontinuity at the edge of the time series becomes significant. 
   The black lines indicate the 95\% confident level. 
   The black points indicate peaks of the wave packets that satisfy the four criteria described in the text.
            }
   \label{fig_wavelet_obs}%
\end{figure}
\subsection{Quasi-periodic-density fluctuations}
Turbulence is the dominant component of density fluctuations in the solar wind, and its characteristic spectra have been observed by radio occultations and in situ observations \citep[e.g.,][]{Spangler1995, Imamura2005, Marsch1990}.
The dissipation of the Alfv\'{e}n waves by turbulence heats the solar wind in the wave- and turbulence-driven model \citep[e.g.,][]{Cranmer2012, Shoda2019}.
However, in this scenario, the density fluctuations associated with MHD waves support the reflection of the Alfv\'{e}n wave, and this reflection plays an important role in the trigger of turbulence. Therefore, there has been interest in studying density fluctuations excited by the Alfv\'{e}n wave, as well as turbulence.

The advection of the plasma density inhomogeneity by the solar wind across the radio ray path causes variations in the frequency/phase of the received signal \citep[e.g.,][]{Efimov2010, Efimov2012, Miyamoto2014}. The phase shift is related to the column density fluctuation, $\delta N_{\mathrm{e}}$, as \citep[e.g.,][]{Imamura2010}
\begin{align}
   \delta \phi & = \frac{a}{c f_0} \delta N_{\mathrm{e}},
\end{align}
where \textit{c} is the speed of light, $f_0$ is the frequency of the radio wave, and $a = e^2 / 8 \pi \varepsilon_0 m_\mathrm{e} \sim 40.3$ m\textsuperscript{3} s\textsuperscript{-2}, with $\varepsilon_0$ being the dielectric constant of the vacuum and $m_\mathrm{e}$ being the electron mass. 

We analyzed the phase time series obtained from the open-loop data by phase unwrapping with a temporal resolution of 0.01 s \citep{Imamura2005, Imamura2010}. 
To suppress noise, we averaged the phase time series in each of the 600-data-point segments to produce a time series with a cadence of 6 s. 
To focus on waves with periods of around 1000 s or shorter, lower-frequency components were excluded by subtracting a fourth-order polynomial fitted to each phase time series. 
Figure \ref{fig_wavelet_obs} shows an example of the electron column density time series processed as above and their wavelet spectra. 
The wavelet spectra were obtained by using the wavelet-transform routine in the Python-based PyCWT ecosystem, which is based on \citet{Torrence1998}. 
The Morlet function was used as the wavelet basis function. 
Spectral peaks are identified as wave packets when the following four criteria are satisfied \citep{Chiba2022}. 
   \begin{itemize}
      \item The peak power exceeds the 95\% confidence level calculated with the method of \citet{Torrence1998}.
      \item The peak is located outside the cone of influence (COI), the region where the effect of the discontinuity at the edge of the time series is significant.
      \item There are no other peaks with higher powers within $\pm$0.5 period in the time and period axis.
      \item The length of the wave packet is longer than the wave period. The packet length is taken to be the full width at half-maximum of the peak in the time axis.
   \end{itemize}
Based on the above criteria, we regard peaks inside the black lines in Figure \ref{fig_wavelet_obs} as significant.

\begin{figure}[htpb]
   \resizebox{\hsize}{!}
   {
    \includegraphics{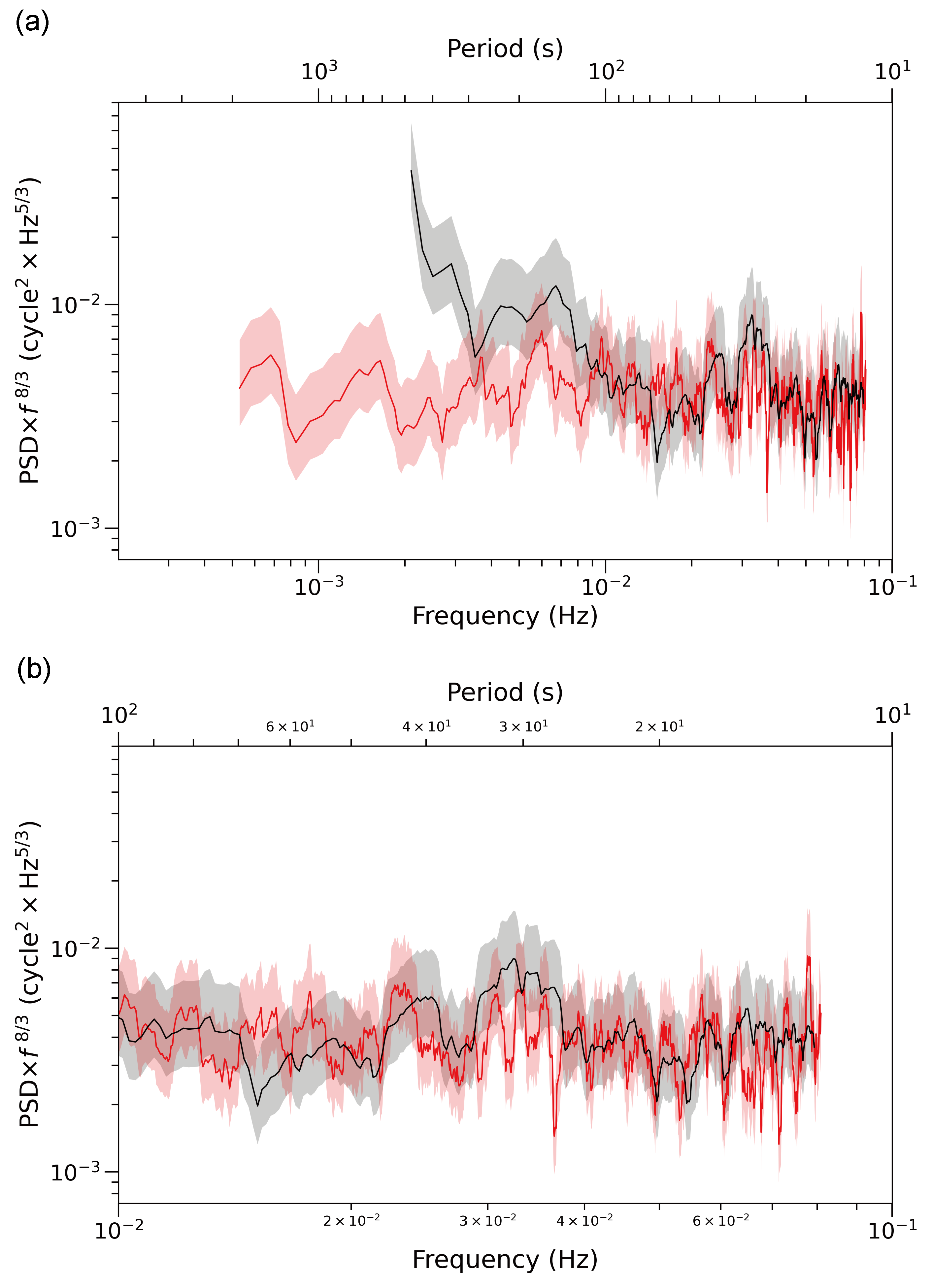}
   }
   \centering
   \caption{(a) Normalized power spectrum of the phase time series data. 
   Red line shows the power spectrum applied to the whole phase time series data of the received signal, and black line shows that the spectrum applied to a part of the time series corresponding to the gray region in the upper panel of Figure \ref{fig_wavelet_obs}. 
   The power spectrum is normalized by multiplying $f^{8/3}$. The 95\% confidence interval of the chi-squared distribution is overplotted as the error bar for each spectrum.
   Panel (b) provides an enlarged view of the spectrum shown in panel (a) for the frequency range of 10\textsuperscript{-2}--10\textsuperscript{-1} Hz.
            }
   \label{fig_PSD}%
\end{figure}

\begin{figure}[htpb]
   \resizebox{\hsize}{!}
   {
   \includegraphics{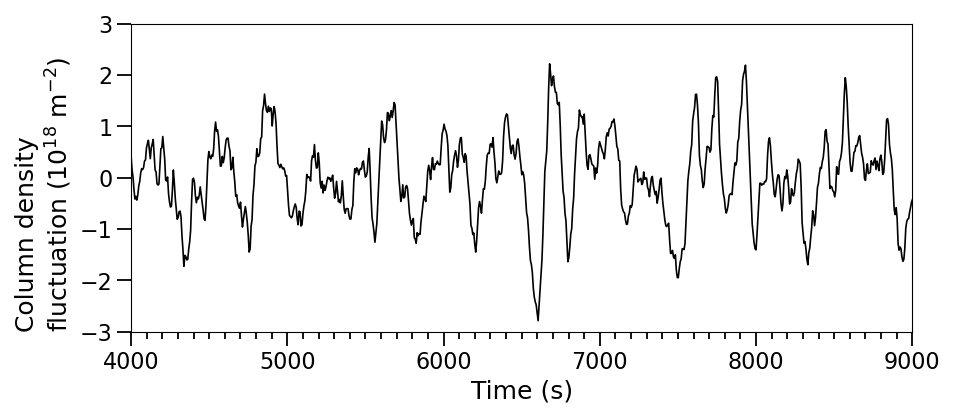}
   }
   \centering
   \caption{Filtered column density fluctuations. The time series corresponds to the gray regions in the upper panel of Figure \ref{fig_wavelet_obs}.
            }
   \label{fig_timeserise}%
\end{figure}

To confirm whether the detected peak is significant, we additionally applied Fast Fourier Transform (FFT) to the phase time series of the received signal corresponding to the upper panel in Figure \ref{fig_wavelet_obs}.
Then, we applied a moving average with a 20-point width kernel to the spectra and calculated the 95\% confidence interval of the chi-squared distribution as the error for each spectrum.
As the power spectrum approximately follows a power-law of $f^{-8/3}$ \citep{Chiba2022}, we normalized the power spectra by multiplying $f^{8/3}$, where $f$ is the temporal frequency. 
Figure \ref{fig_PSD} displays the power spectra applied to the entire data interval and the 5\,000 s interval.
The entire spectrum has peaks with periods of approximately around 1\,000 s, 300 s, and 200 s. 
The spectrum with a 5\,000 s interval additionally has peaks with periods around 200--300 s, and shorter than 100 s. 
These peaks reflect the wave packets detected by the wavelet analysis, as is shown in Figure \ref{fig_wavelet_obs}.
In the upper panel of Figure \ref{fig_wavelet_obs}, the oscillations with periods shorter than 300 s cannot be seen clearly due to small amplitudes and the scale of the panel. 
Figure \ref{fig_timeserise} illustrates the high-pass filtered time series of column density fluctuations (from the gray region in Figure \ref{fig_wavelet_obs}), derived by subtracting a 300-s moving average from the original data. 
Short-period fluctuations are clearly observed in the filtered data.

Most of the wave packets have periods in the range of 20--1\,000 s.
Periods longer than 100 s are ubiquitously detected regardless of the heliocentric distance, being consistent with the previous radio occultation observations by \textit{Ulysses}, \textit{Galileo}, \textit{Mars Express}, \textit{Venus Express}, and \textit{Rosetta} \citep{Efimov2010}. 
In addition to waves with periods longer than $\sim$200 s, which have been identified using the same dataset \citep{Chiba2022}, waves with periods of $\sim$100 s or shorter are also ubiquitously observed at heliocentric distances around 3--6~$R_{\sun}$, 
and these short period components are hardly seen at distances further than 8~$R_{\sun}$.
Such short-period waves have not been studied in previous studies. 
The periods of the detected QPCs seemed to be around 15--5\,000~s, and we discuss the origin of these QPCs by comparing the radio occultation data with the MHD simulation in the below sections.

%
%
\section{Density fluctuations in a three-dimensional \\ MHD simulation}
Radio-wave observations presented in the previous section show the presence of quasi-periodic density fluctuations in the solar wind. 
To understand the physical characteristics of the observed fluctuations, the observed quasi-periodic density fluctuations given in the previous section are compared with the density fluctuations in the MHD simulation presented by \citet{Shoda2021}. 
The four-dimensional data generated by the model in their study are used in the analysis below.
\subsection{Simulation data}
The numerical data analyzed in this study is retrieved from a three-dimensional MHD simulation of the wave- and turbulence-driven model of the solar wind \citep{Shoda2021}. 
In this simulation, an Alfv\'{e}n-wave driver is imposed at the base of the corona and the wave propagation and wind acceleration are numerically calculated in a radially diverging flux tube. 
The Alfv\'{e}nic slow solar wind observed by the first encounter of PSP is reproduced, with a small but non-zero fraction of magnetic switchbacks within the simulation domain. The detailed numerical setup and the simulation results are presented in \citet{Shoda2021}. 
We analyze data for a duration of 12104 s after the system reached a quasi-steady state. The time resolution of the numerical data is 6 s.

To directly compare the simulation result with the observations, we calculate the column density integrated along the line of sight. However, the simulation domain is insufficient to provide a sufficiently long line of sight. 
To resolve this issue, the simulation domain is periodically extended in the azimuthal direction using the periodic boundary condition applied in the simulation.
Specifically, we first define a virtual line of sight at a specified heliocentric distance ($R$).
Using $s$ as the coordinate along this line, we proceed with the following integration.
\begin{align}
    N_\mathrm{e} (t, R) = \int_{-l_{\rm LOS}/2}^{l_{\rm LOS}/2} n_{\mathrm{e}} \Big(t, r_{\rm LOS}(s,R), \theta_{\rm LOS}, \phi_{\rm LOS}(s,R) \Big) \ ds, \label{eq:column_density_definition}
\end{align}
where $n_e(t, r, \theta, \phi)$ is the electron number density in the spherical coordinate (used in the simulation) and
\begin{align}
    r_{\rm LOS}(s,R) = \sqrt{s^2+R^2}, \ \ \theta_{\rm LOS}=0, \ \ \phi_{\rm LOS}(s,R) = \arctan (s/R).
\end{align}
To reduce the computational cost without loss of accuracy, we set the length of the line of sight as a function of $R$:
\begin{align}
    l_{\rm LOS} = 4 \pi R.
\end{align}
To compute the integral in Eq.~\eqref{eq:column_density_definition}, we discretize $s$ according to the following relation:
\begin{align}
    \Delta s = 2 R \Delta \phi,
\end{align}
where $\Delta \phi = 2.0 \times 10^{-4} {\rm \ rad}$ is the resolution of the simulation data in the azimuthal direction.
We confirmed that $l_{\rm LOS}$ is sufficiently large and $\Delta s$ is sufficiently small, ensuring that $N_e$ does not depend on the selection of these values.
\begin{figure*}[htpb]
   \includegraphics[width=17cm]{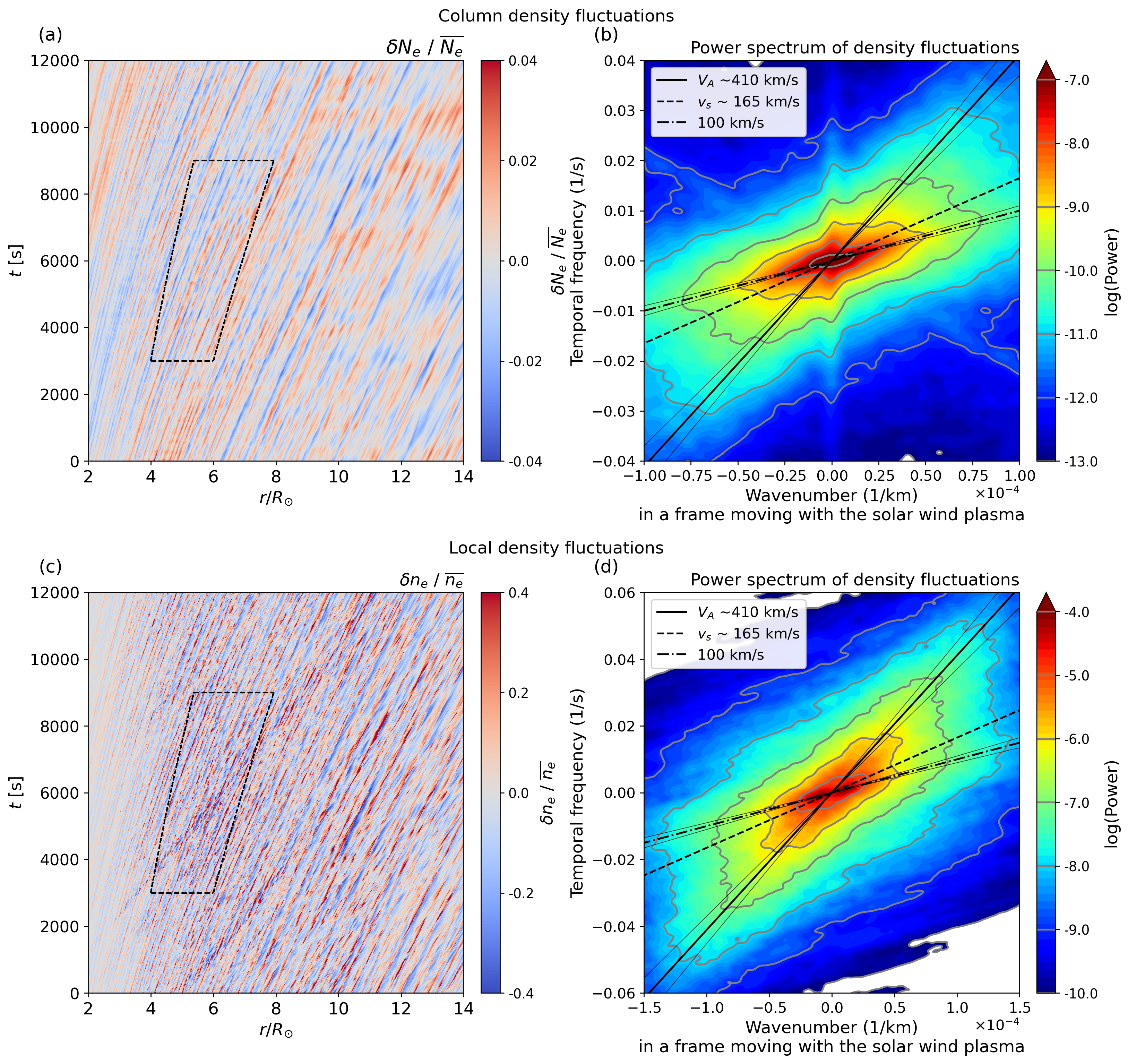}
   \centering
   \caption{
   Panel (a), (c): Distance-time diagram of fractional density fluctuations in column density, $\delta N_e/ \overline{N_e}$, (upper panel), and local density, $\delta n_e/ \overline{n_e}$, (lower panel), within the MHD simulation. The area encircled by dashed black lines is designated for subsequent Fourier analysis. 
   Panel (b), (d): Two-dimensional power spectrum of density fluctuation with respect to wavenumber and frequency. The spectra are calculated in a coordinate system aligned with the background solar wind. Gray lines represent power contours from the FFT spectrum. Thick solid and dashed lines denote nominal Alfv\'en and sound speeds, respectively, while a dash-dotted line marks 100 km/s. Thin lines outline the phase speed range useful for mode decomposition.
            }
   \label{fig_model_FFT}%
\end{figure*}

\subsection{Spectral analysis of density fluctuations in the model}
Figure \ref{fig_model_FFT}(a) and (c) show the radial distance-time (\textit{r-t}) diagram of the fractional fluctuations of the column density, $\delta N_\mathrm{e}/ \overline{N_\mathrm{e}}$, and that of the local density, $\delta n_\mathrm{e}/ \overline{n_\mathrm{e}}$, where $\delta N_\mathrm{e}$ is the column density fluctuation, $\overline{N_\mathrm{e}}$ is the time average of the column density at each location, and those of the density fluctuation correspond to $\delta n_\mathrm{e}$ and $\overline{n_\mathrm{e}}$.
We use the local density at $\theta = 0$ and $\varphi = 0$. Here, we use the time average of the density fluctuation as the background density.
The dominance of outwardly propagating structures is evident.
The \textit{r-t} diagram of the local density shows finer structures than the column density diagram. 
This is probably because the integration along the ray path suppresses local density fluctuations. 
Furthermore, a superposition of two wave-like structures with different slopes (phase speeds) is noticeable around 2–8~$R_{\sun}$, in particular in the local density map. 
The characteristics of these waves are further analyzed in the following section.

The observed phase speeds are modulated by the background solar wind, and thus we remove this effect as follows. 
First, the radial distribution of the density fluctuation around 4--6~$R_{\sun}$ is sampled along trajectories, corresponding to the coordinate system moving at a background velocity of each point \citep{Shoda2021}; 
the range of the sampled data is shown by the black surrounded region in Figure \ref{fig_model_FFT}(a) and (c). 
Applying the discrete Fourier transform (DFT) with the Hann window and Gaussian smoothing, the two-dimensional spectrum with respect to frequency and wavenumber of the density fluctuation is obtained.
The spectrum in Figure \ref{fig_model_FFT}(b) and (d) appears to be composed of two components with different phase velocities. 
The faster one corresponds to the mean Alfv\'{e}n speed, $V_{\mathrm{A}}$, which is approximately 410 km s\textsuperscript{-1} in the analyzed region, and is considered to represent fast magnetoacoustic waves. 
The slower ones correspond to approximately 100 km s\textsuperscript{-1}, which is slower than the sound speed of $v_{\mathrm{S}} =$ 165 km s\textsuperscript{-1}, and are considered slow magnetoacoustic waves.
Propagating diagonally relative to the radial magnetic field, the propagation speed decreases against the radial direction. 
Hence, the peak with the phase speed of 100~km~s\textsuperscript{-1} can be interpreted as slow magnetoacoustic waves.
Slow magnetoacoustic waves have a stronger peak at low wavenumbers ($\la$10\textsuperscript{-5} km\textsuperscript{-1}) in both the column and local density, and both the fast and slow magnetoacoustic waves extend to $\sim$10\textsuperscript{-4} km\textsuperscript{-1}. 
We note that Alfv\'{e}n waves are incompressible and are not accompanied by density fluctuations that we analyze.

\begin{figure*}[htpb]
   \includegraphics[width=17cm]{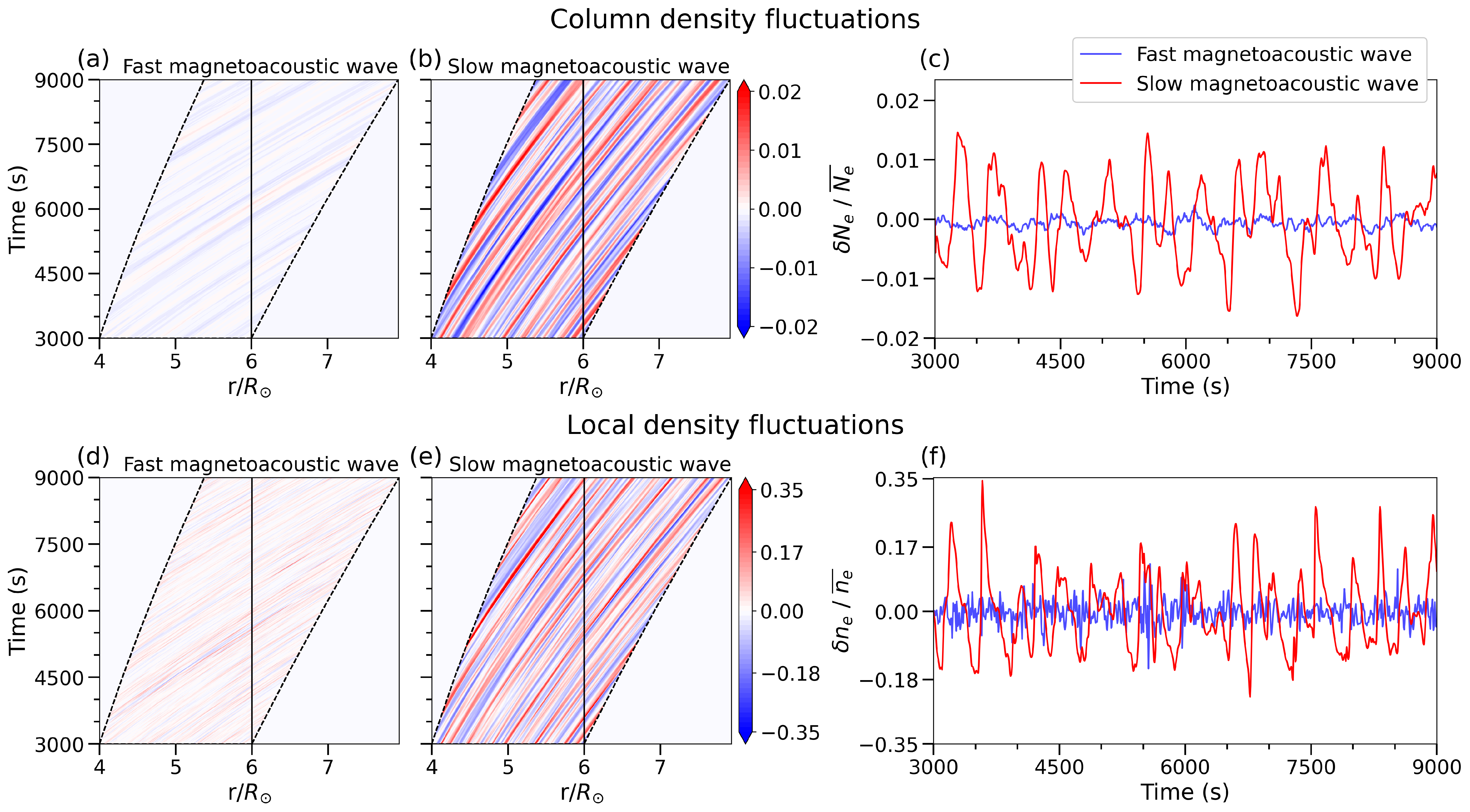}
   \centering
   \caption{
   Filtered \textit{r-t} diagram for (a), (d) fast magnetoacoustic waves and that for (b), (e) slow magnetoacoustic waves. 
   The region encircled by dashed black lines in panels (a)/(b) and (d)/(e) correspond to those in Figure \ref{fig_model_FFT}(a) and (c), respectively.
   Panel (c), (f): Time series of the density perturbation at a heliocentric distance of 6~$R_{\sun}$ filtered to retain fast (blue line) and slow (red line) modes.
   These time series correspond to the solid lines in panels (a) and (b) and are used in the subsequent analysis (Figure \ref{fig_wavelet_model}). 
   The upper panels, (a)--(c), are obtained from the fractional fluctuations of the column density, and the lower panels, (d)--(f), are obtained from those of the local density. 
            }
   \label{fig_decomposed}%
\end{figure*}

To decompose the density fluctuations into fast and slow magnetoacoustic waves, the inverse DFT was applied to the complex amplitudes in the hourglass-shaped areas bounded by the phase velocities of 410 $\pm$ 41.0 km s\textsuperscript{-1} for fast magnetoacoustic waves and 100 $\pm$10.0 km s\textsuperscript{-1} for slow magnetoacoustic waves, as is shown in Figure \ref{fig_model_FFT}(b) and (d). 
We note that we used the raw spectrum, without convolution of the window function and smoothing, for the wave decomposition.
This procedure yields a filtered \textit{r-t} diagram for each of the fast and slow magnetoacoustic waves in the coordinate system moving with the background velocity. 
The range of the phase velocity was determined on the assumption that the velocity can be variable by $\pm$10\%. The results are almost unchanged when we used $\pm$5\% and $\pm$20\% for the range of the phase velocity. 
The filtered \textit{r-t} diagrams were re-arranged onto the original coordinate system to compare the result with the radio occultation measurements (Figure \ref{fig_decomposed}(a), (b), (d), and (e)). 
The time series of the density perturbation at 6~$R_{\sun}$ was extracted from the \textit{r-t} diagram for each of the fast and slow magnetoacoustic waves, as is shown in Figure \ref{fig_decomposed}(c) and (f). 
According to these Figures, slow magnetoacoustic waves tend to have larger amplitudes and longer periods than fast magnetoacoustic waves; this is consistent with the fact that the spectrum of slow magnetoacoustic waves in Figure \ref{fig_model_FFT}(b) and (d) have a stronger peak at the low wavenumber ($\la$10\textsuperscript{-5} km\textsuperscript{-1}). 
Figures \ref{fig_decomposed}(c) and (f) show that the amplitude of the local density is around ten percent, which is comparable to the observed values derived in \citet{Chiba2022}, while that of the column density is around several percent.

\begin{figure*}[htpb]
   \includegraphics[width=17cm]{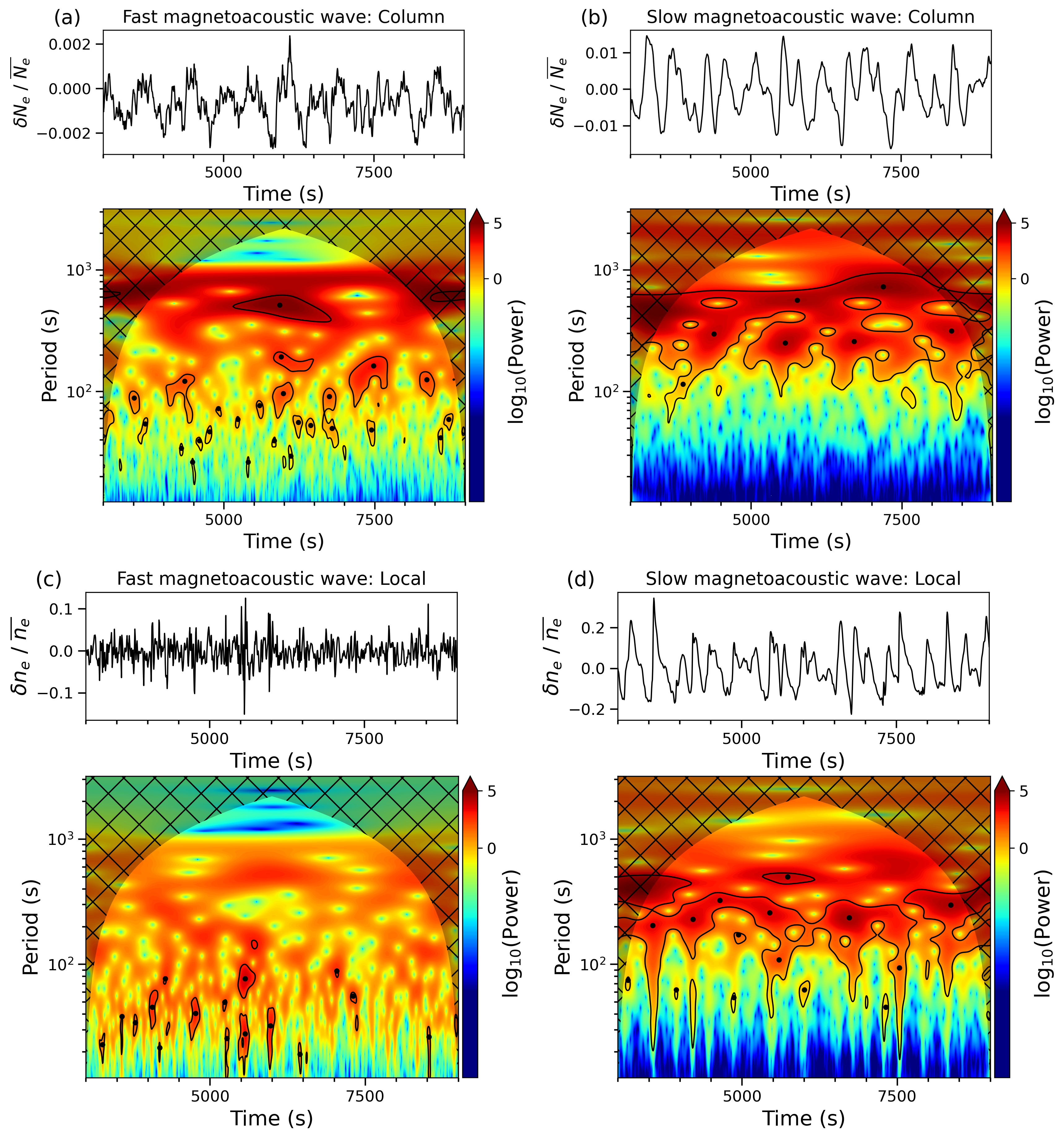}
   \centering
   \caption{Time series of the fractional density perturbation at 6~$R_{\sun}$ in the model (upper panel in each of a,b,c,d) and its wavelet spectrum (lower panel in each of a,b,c,d), for the (a)/(c) fast and (b)/(d) slow magnetoacoustic waves extracted from the original time series. The wavelet power is normalized by the variance of the time series. The shaded region indicates the COI. The black lines indicate the 95\% confidence level. The black points indicate peaks of the wave packets that satisfy the four criteria described in the text. The data in panels (a) and (b) are decomposed by the fractional fluctuations of the column density, and the data in panels (c) and (d) are decomposed by the fractional fluctuations of the local density.
            }
   \label{fig_wavelet_model}%
\end{figure*}

To compare the wave characteristics in the model with those observed by the radio occultation, we applied wavelet transform to the filtered time series at the same distances as the observations. 
Figure \ref{fig_wavelet_model} shows samples of the obtained wavelet spectra from the filtered time series at 6~$R_{\sun}$.
In the wavelet spectra of the column density fluctuations, fast magnetoacoustic waves are found at periods from $\sim$20 to 500 s, and slow magnetoacoustic waves are found mostly around periods of 100--800 s.
Meanwhile, in the wavelet spectra of the local density fluctuations, fast magnetoacoustic waves are found at periods shorter than 100 s, and slow magnetoacoustic waves are found mostly around periods of 100--500 with a small fraction at < 100 s.
Although the fast magnetoacoustic waves in Figure \ref{fig_wavelet_model}(a) appear to have a broadened distribution, in Figure \ref{fig_decomposed}(c) and (f), the column density fluctuations exhibit approximately ten times smaller amplitudes of the fast modes than those of the slow modes, and the fast modes in the local density fluctuations have comparable amplitudes to the slow modes.
Figure \ref{fig_model_FFT} and \ref{fig_decomposed} suggest that fine structures in the local density were smoothed out in the column density by the integration along the ray path. 
Since the component of the fast modes is not distinct, there is a possibility that, in the column density fluctuations, the mode decomposition did not appear to succeed well, and the fast modes in the column density detected may include the pseudo-long-time variations. 

The wave periods obtained from the radio occultation data (Figure \ref{fig_wavelet_obs}) approximately range from 15 to 5\,000 s. 
This range appears to be more extended than the period distribution of either the fast or slow magnetoacoustic waves suggested by the model.
Some of the wave packets observed have periods longer than the MHD simulation.
This is because the wavelet analyses shown above limit the extracted time series as 6\,000 s.
Focusing on the wave packets with periods shorter than 1\,000 s, the wave-period analysis presented here indicates that the observed density fluctuations appear to be a superposition of fast and slow magnetoacoustic waves.
Therefore, to explain the extended wave-period distribution in the observation, both fast and slow magnetoacoustic waves are considered to be required.

\section{Wave statistics}
\begin{figure*}[htpb]
    \sidecaption
   \includegraphics[width=12cm]{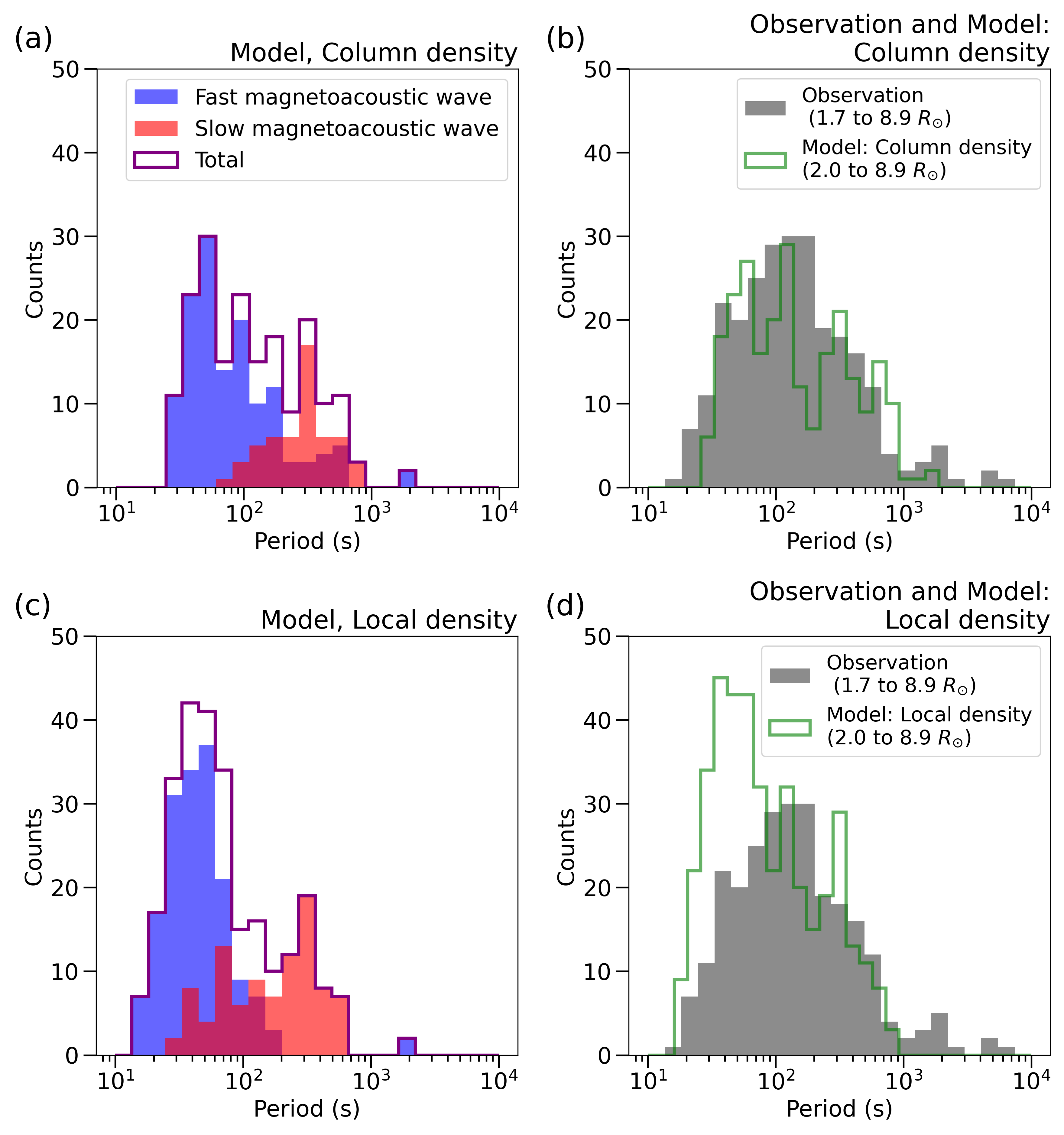}
   \centering
   \caption{
   Histograms of the wave period detected from the decomposed fractional fluctuations of (a) the column density and (c) local density. 
   In panels (a) and (c), the counts for fast magnetoacoustic waves and slow magnetoacoustic waves are shown separately in addition to the total counts.
   Panel (b), (d): Comparison of the period histogram of the density fluctuations detected from the observations and the entire time series of the model.
   Panels (b) and (d) used the column density fluctuations and the local density fluctuations as the model data, respectively.
   In panels (b) and (b), the observations are displayed in gray bins, and the models are shown in green bins. 
   We use the observational data in the heliocentric distance range of 1.7--8.9~$R_{\sun}$ and the model data in the range of 2.0--8.9~$R_{\sun}$.
   The bin size is such that the whole range of 10\textsuperscript{1}--10\textsuperscript{4} is divided into 30 regular subintervals on a logarithmic scale.
            }
   \label{fig_histogram}%
\end{figure*}
In this section, we investigate the detailed characteristics of the density fluctuations.
First, we compare the wavelet spectra of the model with the observation.
Figures \ref{fig_histogram}(a) show the histograms of the wave packet's period for fast and slow magnetoacoustic waves detected in the column and local density fluctuations in the model. 
The fast and slow magnetoacoustic waves obtained from the column density in the model mostly have periods of \\ 20--700 s and 60--1\,000 s, respectively, and those obtained from the local density mostly have periods of 20--200 s and 30--700 s, respectively.
Since the fast modes in the column density may have artificial components, as was mentioned in the previous section, the fast modes in the column density have a more broadened histogram than those in the local density.
The total histogram has a broadened distribution due to the superposition of the fast and slow magnetoacoustic waves in both the column and local density. 
Figures \ref{fig_histogram}(b) compare the histograms of the wave period for all observations with the model. 
In Figures \ref{fig_histogram}(b), we use the whole time series of the column and local density fluctuations at each point as “Model” data and compare them with the observations.
The histograms for individual observations are given in the Appendix. The wave periods are distributed from $\sim$20 to $\sim$7\,000 s with a peak around 100 s.
\\
Comparing the local density fluctuations in the model and the observation, waves with periods longer than 1\,000 s are almost absent in the model. 
This discrepancy is possibly due to the artificial nature of the wave driver in the simulation. 
In the model, Alfv\'{e}n waves are forced to exhibit peak energy at periods of 1\,000 s, while observed Alfv\'{e}n waves have longer periods.
This results in the underestimation of long-period waves in the model. 
Given the similarity of the histogram between the model and the observation, it is probable that the observed QPCs also include fast magnetoacoustic waves at short periods and slow magnetoacoustic waves at long periods with significant overlap.

Next, based on the method of \citet{Chiba2022}, we calculated the column density amplitude of each observed wave packet as $N_\mathrm{e}' = (cf_0 / a) \, \phi'$, where $\phi'$ is the phase amplitude of each wave packet, and $\tau_\mathrm{packet}$ is the duration of the wave packet. 
Furthermore, the local density amplitude, $n_\mathrm{e}'$, was calculated. 
Here, the prime means the amplitude of each quantity.
To convert $N_\mathrm{e}'$ to $n_\mathrm{e}'$, the spatial scale of the density fluctuation along the line of sight is assumed in two different ways.

The first assumption (assumption 1) is that the spatial scale of the density fluctuation is the same as the radial length of the wave packet, $n_\mathrm{e}' = N_\mathrm{e}' / \tau_\mathrm{packet} (v_\mathrm{group} + V_\mathrm{SW})$, where $v_\mathrm{group}$ is the group velocity of the wave, and $V_\mathrm{SW}$ is the solar wind velocity of the background.
$V_\mathrm{SW}$ is taken from \citet{Chiba2022}, and $v_\mathrm{group}$ is taken to the sound speed, $v_\mathrm{S}$, under the assumption that the fluctuations are dominated by outwardly propagating slow magnetoacoustic waves. 
Another assumption (assumption 2) is that the density fluctuation extends the order of several $R$ ($\sim \pi R$) along the line of sight.
In this assumption, the spatial scale is estimated by the equation of the column density,
\begin{align}
   N_\mathrm{e}' & = \int dS \ n_\mathrm{e}'(r).
\end{align}
$dS$ is the integration path increment of the line of sight.
This assumption is based on the effective thickness of the background plasma density along the line of sight.
Using these local density amplitudes, $n_\mathrm{e}'$, the fractional density amplitude, $n_\mathrm{e}' / n_0$, the ratio of the local density fluctuations to the background electron density, is obtained by adopting $n_0(r)$ from the empirical model given by \citet{Wexler2019b}.

\begin{figure*}[htpb]
   \sidecaption
   \includegraphics[width=12cm]{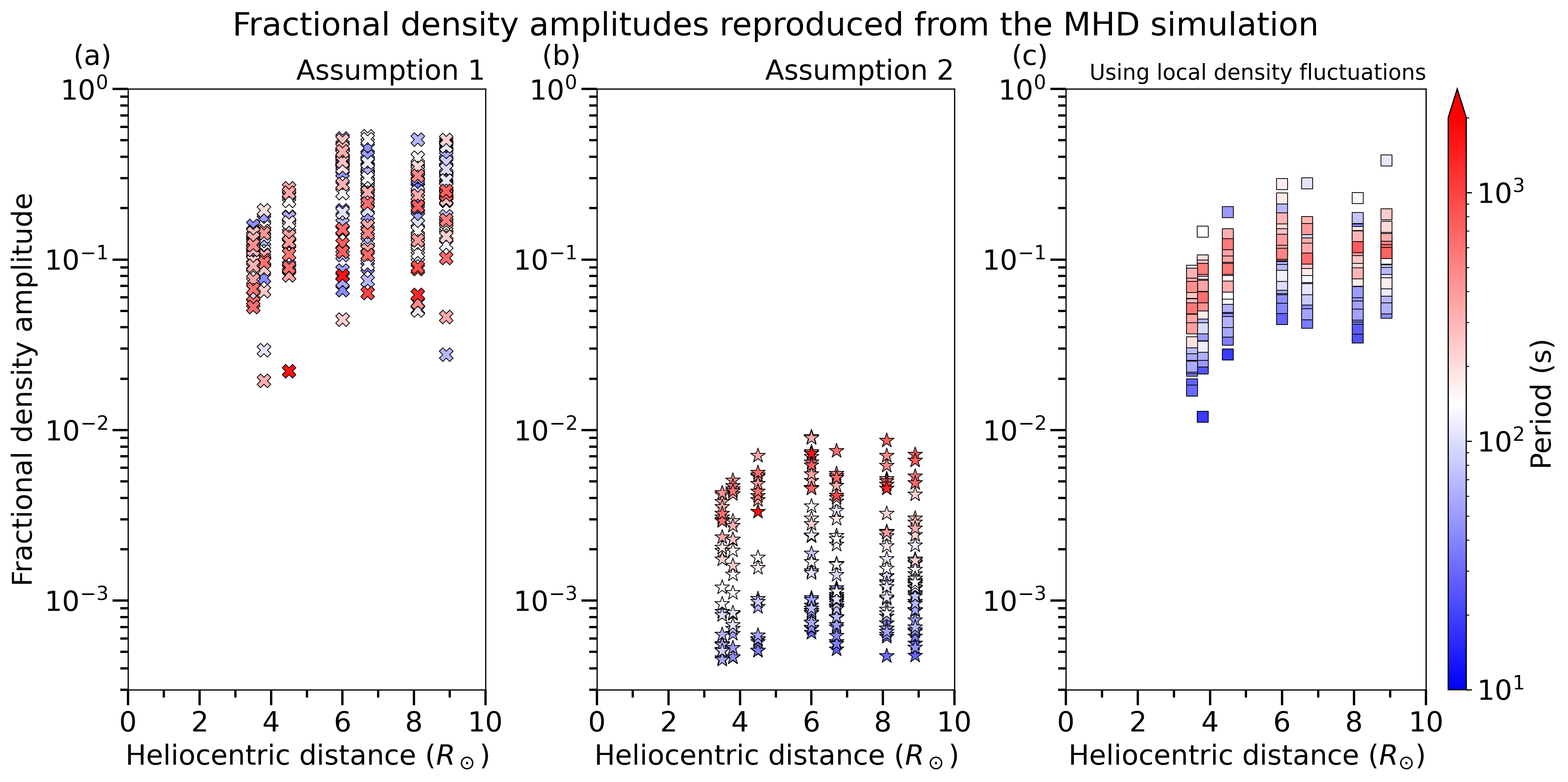}
   \centering
   \caption{Radial distributions of the fractional density amplitude, $n_\mathrm{e}' / n_0$, of the detected wave packets obtained from the model. 
   Panel (a) shows the assumption: the scale of the wave packets is the same as the radial length of the wave packet (crosses). Panel (b) indicates the assumption: the density fluctuation extends the order of several $R_{\sun}$ along the line of sight (stars). 
   Panel (c) shows the amplitudes, which are directly calculated from the local density fluctuations (squares). 
   The color represents the period of each wave packet.
            }
   \label{fig_fractional_model}%
\end{figure*}

Figure \ref{fig_fractional_model} compares the fractional density amplitudes calculated from the column density reproduced by the model with those directly calculated from the local density.
According to this Figure, assumption 1 and the local density fluctuations have larger amplitudes than those of assumption 2.
Since the effective thickness of the background plasma density is much larger than the spatial scale of the MHD waves along the radial direction, the density fluctuations are elongated along the line of sight in assumption 2. For this reason, assumption 2 underestimates the spatial scale of the wave packets.
Therefore, we adopt assumption 1 to calculate the fractional density amplitudes from the observational data.

\begin{figure}[htpb]
\resizebox{\hsize}{!}
   {
   \includegraphics[width=12cm]{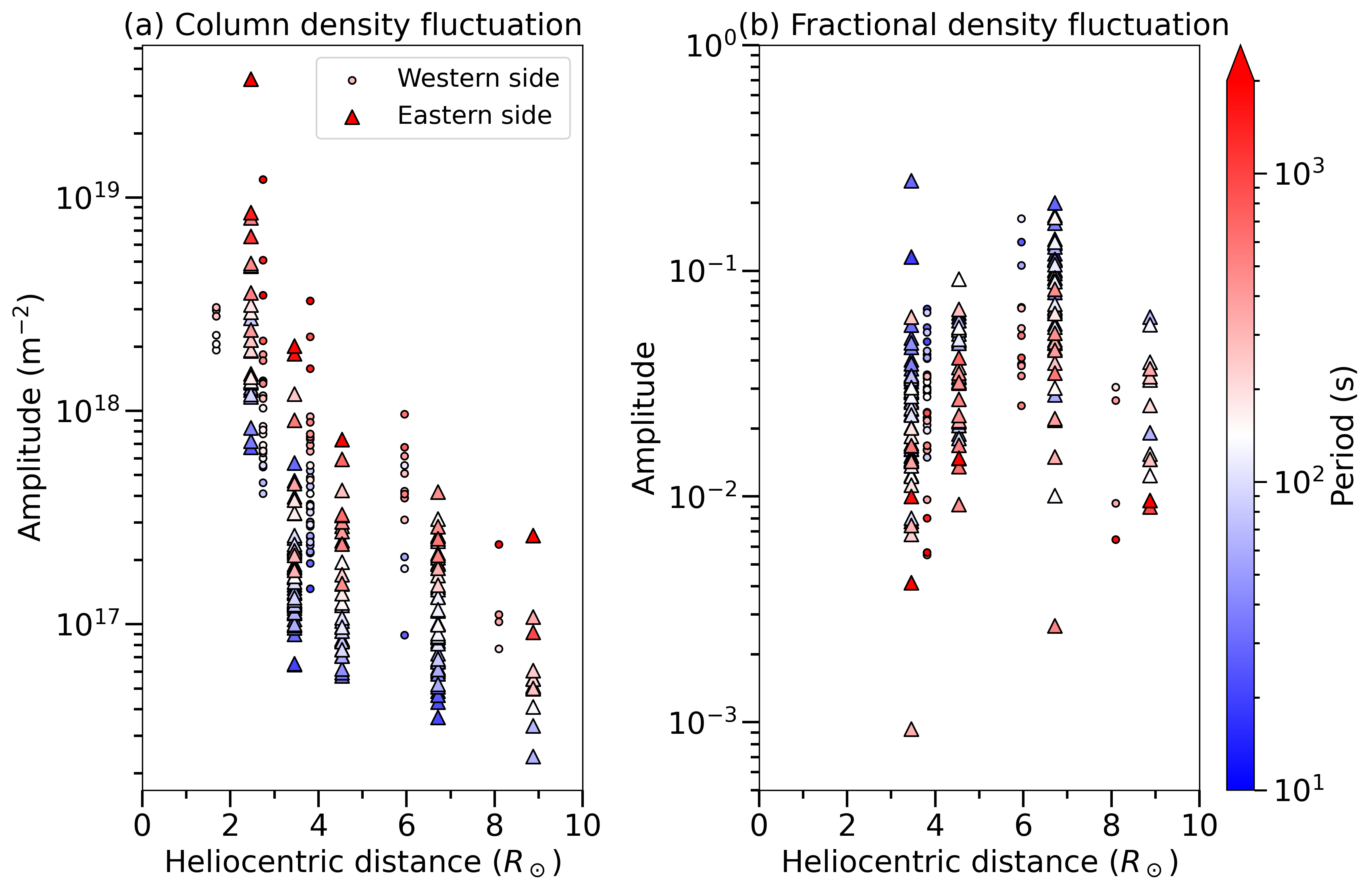}
   }
   \centering
   \caption{Radial distributions of the amplitude of the detected wave packets obtained from the observations.
   Panel (a) shows the amplitude of the electron column density fluctuation, $N_\mathrm{e}'$,  and panel (b) shows that of the fractional density fluctuation.
   Circles show observations on the western side of the Sun and triangles show observations on the eastern side. The color represents the period of each wave packet.
            }
   \label{fig_column_obs}%
\end{figure}

Figure \ref{fig_column_obs} shows the radial distributions of $N_\mathrm{e}'$, and that of $n_\mathrm{e}' / n_0$.
In Figure \ref{fig_column_obs}(b), the local density amplitudes at distances closer than 3~$R_{\sun}$ were not estimated because the solar wind velocity was derived from the data further than 3~$R_{\sun}$ only due to the strong scattering at shorter distances \citep[see][Section 3]{Chiba2022}. 
The results are qualitatively similar to those in \citet{Chiba2022}, with more contributions from short-period waves due to the improvement of the analysis. 
In Figure \ref{fig_column_obs}, the maximum value of the $n_\mathrm{e}' / n_0$ is on order of 10\%. 
The $n_\mathrm{e}' / n_0$ peaks around 6~$R_{\sun}$ and decreases beyond 8~$R_{\sun}$, suggesting the dissipation of slow magnetoacoustic waves at >8~$R_{\sun}$. 
Long-period waves tend to have larger amplitudes of $N_\mathrm{e}'$ than short-period waves, while short-period waves have greater amplitudes of $n_\mathrm{e}' / n_0$ than long-period waves. 
This is because a longer period means a larger spatial scale of the wave packet, resulting in a longer integration of the wave amplitude along the ray path.

\begin{figure}[htpb]
\resizebox{\hsize}{!}
{
\includegraphics{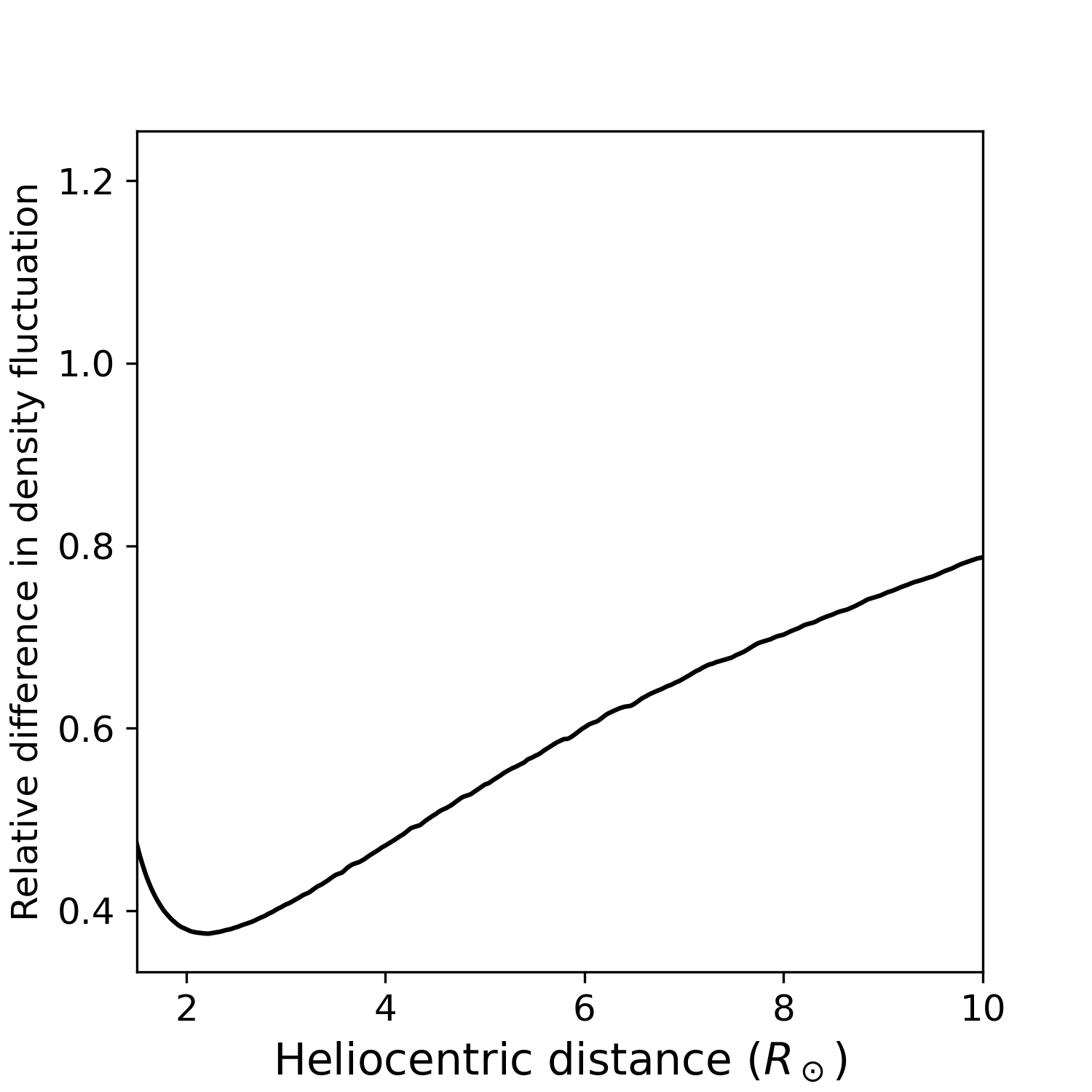}
}
   \centering
   \caption{Relative difference in the estimated density amplitude between the case where the group velocity is the sound speed, $v_\mathrm{S}$, and the case where it is the Alfv\'{e}n speed, $V_\mathrm{A}$. The $v_\mathrm{S}$ was taken to be 165 km s\textsuperscript{-1} for the temperature of 10\textsuperscript{6} K, and $V_\mathrm{A}$ was obtained by the model.
            }
   \label{fig_velocity_ratio}%
\end{figure}

If the detected waves are fast magnetoacoustic waves, the use of $v_\mathrm{S}$ for $v_\mathrm{group}$ would cause errors in the estimation of $n_\mathrm{e}'$. The relative change of this estimation associated with the replacement of $v_{\mathrm{S}}$ with $V_{\mathrm{A}}$ is shown in Figure \ref{fig_velocity_ratio}, suggesting that will be decreased by about 20--60\% for fast magnetoacoustic waves. Since the $n_\mathrm{e}'$ amplitudes of short-period (10--100 s) waves seem to be 2--4 times larger than that of long-period (100--1\,000 s) waves (Figure \ref{fig_column_obs}(b)), the correction for fast magnetoacoustic waves above may obscure the dependence of the amplitude on the wave period if the short-period waves are mostly fast magnetoacoustic waves and the long-period waves are slow magnetoacoustic waves. Regardless of the density fluctuation mode, the observed shorter period waves show density fluctuations comparable to or larger than longer waves, contrasting with the simulation where fluctuations are larger for longer periods.

\section{Summary}
The density oscillations observed in the solar corona by radio occultation have been attributed to slow magnetoacoustic waves in previous studies \citep[e.g.,][]{Efimov2010, Miyamoto2014}; however, recent in situ observations and numerical models indicated the existence of various types of waves \citep[e.g.,][]{Chaston2020, Suzuki2005}. 
To investigate the nature of the density oscillation, we compare the radio occultation data taken in 2016 using the JAXA’s Venus orbiter \textit{Akatsuki} with the MHD simulation conducted by \citet{Shoda2021}. 
The time-spatial spectrum of the density fluctuation in the model exhibits two components that are considered to be fast and slow magnetoacoustic waves. 
The fast magnetoacoustic waves in the model tend to have periods shorter than the slow magnetoacoustic waves, and the superposition of these modes has a broadened histogram of the period extending in the range of 10--500~s, which resembles the histogram for the observed waves. 
Based on this comparison, it is probable that the density oscillations observed by radio occultation also include fast and slow magnetoacoustic waves, and that fast magnetoacoustic waves are predominant at short periods and slow magnetoacoustic waves are prevalent at long periods. 
The observed density amplitudes of short-period waves are comparable to or larger than those of long-period waves. 
On the other hand, long-period waves tend to have larger amplitudes of the column electron density integrated along the ray path, which is directly observed by radio occultation, than short-period waves, because of the longer integration paths for long-period waves. 

To better constrain the wave modes, observations of the magnetic field oscillation and density oscillation are needed. Magnetic field oscillations can be observed through the Faraday rotation (FR), which can be obtained from the phase shift between right- and left-hand circular polarization waves in radio occultation experiments \citep{Patzold1987, Jensen2013a, Jensen2013b, Wexler2019a}. 
Those circular polarized waves were recorded in the experiment used in this study. 
A more detailed comparison between fast and slow magnetoacoustic waves with this technique is left for future studies. 

In recent years, PSP has directly observed solar corona and provided in situ data on electric and magnetic fields, density, waves, and particles. The results from the PSP’s first perihelion passage at distances from 53 to 35~$R_{\sun}$ enabled the spectral composition of MHD waves \cite{Chaston2020}. It was shown that slow magnetoacoustic waves were more prominent than fast magnetoacoustic waves at periods longer than 100 s, and the MHD waves with periods shorter than 100 s contained fast and slow magnetoacoustic waves with similar magnitudes. These results are qualitatively consistent with our conclusion that fast and slow magnetoacoustic waves coexist and that slow magnetoacoustic waves are more prominent at longer periods. Although the radial distances observed by radio occultations and in situ observations differ, combining these two methods, together with optical observations, will enable us to derive a radial distribution of physical parameters and trace energy transport in the whole acceleration region. 

\begin{acknowledgements}
We thank the \textit{Akatsuki} project team, the VLBI group, and the UDSC operation team for providing valuable suggestions. 
The simulation data used in this study were calculated on the Cray XC50 at the Center for Computational Astrophysics, National Astronomical Observatory of Japan.
S.C. was supported by a Grant-in-Aid for Japan JSP Fellows and JST SPRING, Grant Number JP23KJ0481 and JPMJSP2108.
M.S. is supported by JSPS KAKENHI Grant Number JP22K14077.
\end{acknowledgements}

%
\bibliographystyle{bibtex/aa} 
\bibliography{bibtex/references} 
%

%
\begin{appendix}
\section{}    
Supplemental figures mentioned in Section 4 are given.
\begin{figure}[htpb]
  \includegraphics[width=0.95\linewidth]{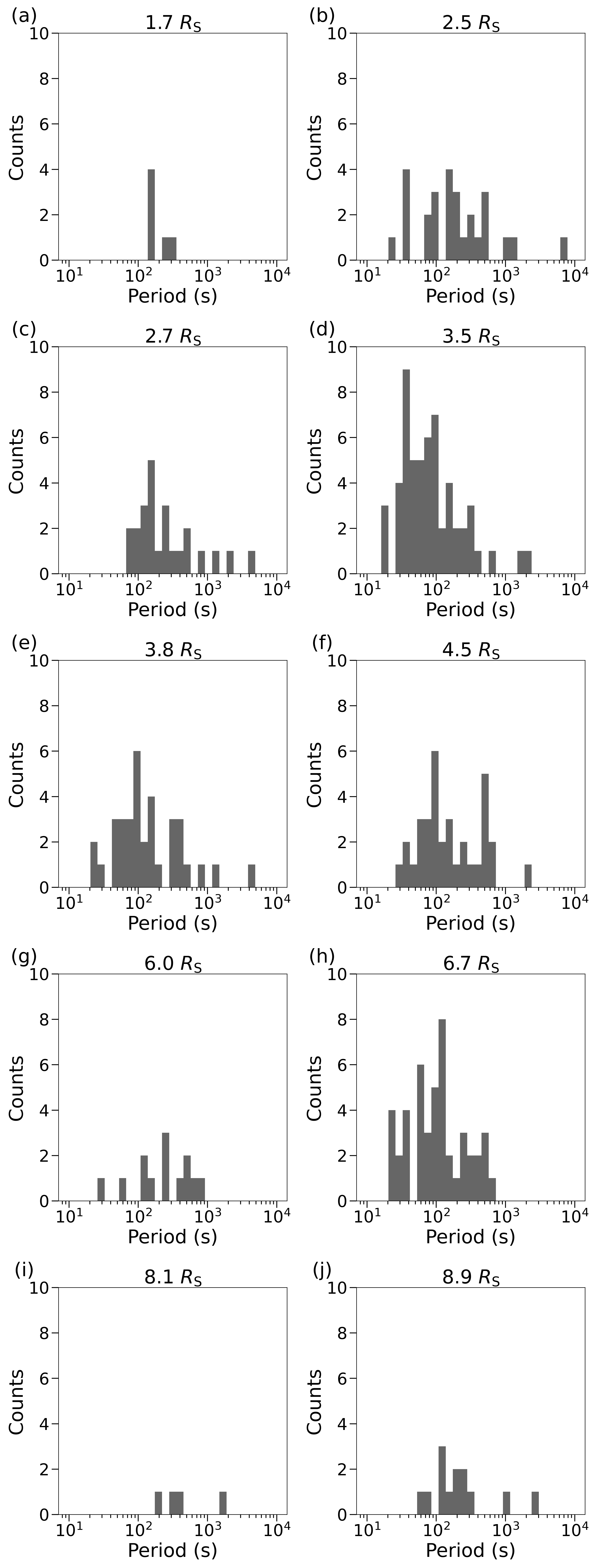}
   \centering
   \caption{Histograms of the wave period for the individual observations (a)–(j) in the heliocentric distance range of 1.7--8.9~$R_{\sun}$. The manner of determining the bin size is the same as in Figure \ref{fig_histogram}(b).
            }
   \label{Fig_app}%
\end{figure}
\end{appendix}
\end{document}